\documentclass[trackchanges,twocolumn]{aastex7}

\usepackage{amsmath}
\usepackage{graphicx}
\usepackage{caption}
\usepackage{subcaption}
\usepackage{hyperref}
\newcommand{\Sg}{Sgr~A*}
\newcommand{\Ms}{Mini-spiral}
\newcommand{\0}{\phantom{0}}
\newcommand{\kms}{\hbox{km~s$^{-1}$}}

\begin{document}

\title{Mid-infrared extinction toward the Galactic center}

\author[orcid=0000-0000-0000-0001,sname='North America']{Sebastiano D. von Fellenberg}
\altaffiliation{Feodor Lynen Fellow}
\affiliation{Canadian Institute for Theoretical Astrophysics, University of Toronto, 60 St. George Street, Toronto, ON M5S 3H8, Canada}
\affiliation{Max Planck Insitute for Radioastronomy, auf dem H{\"u}gel 69, Bonn, Germany}
\email[show]{sfellenberg@cita.utoronto.ca}  

\author[orcid=0000-0003-3503-3446,sname=Michail, gname=Joseph]{Joseph M. Michail}
\altaffiliation{NSF Astronomy \& Astrophysics Postdoctoral Fellow}
\affiliation{Center for Astrophysics $|$ Harvard \& Smithsonian, 60 Garden Street, Cambridge, MA 02138, USA}
\email[show]{joseph.michail@cfa.harvard.edu}

\author[0000-0002-9895-5758]{S. P. Willner}
\affiliation{Center for Astrophysics $|$ Harvard \& Smithsonian, 60 Garden Street, Cambridge, MA 02138, USA}
\email{swillner@cfa.harvard.edu}

\author[orcid=0009-0001-1040-4784]{Braden Seefeldt-Gail}
\affiliation{Canadian Institute for Theoretical Astrophysics, University of Toronto, 60 St. George Street, Toronto, ON M5S 3H8, Canada}
\email{braden.gail@mail.utoronto.ca}
\affiliation{Dunlap Institute for Astronomy and Astrophysics, University of Toronto, 50 St. George Street, Toronto, ON M5S 3H4, Canada}
\affiliation{Dunlap Institute for Astronomy and Astrophysics, University of Toronto, 50 St. George Street, Toronto, ON M5S 3H4, Canada}

\author[0009-0003-9906-2745]{Tamojeet Roychowdhury}
\affiliation{Department of Astronomy, University of California Berkeley, Berkeley, CA 94704, USA}
\email{tamojeet@iitb.ac.in}

\author[0000-0003-4801-0489]{Macarena Garcia Marin}
\affiliation{European Space Agency (ESA), ESA Office, Space Telescope Science Institute, 3700 San Martin Drive, Baltimore, MD 21218, USA}
\email[]{Macarena.Garcia.Marin@esa.int}

\author[0000-0002-0670-0708]{Giovanni G. Fazio}
\affiliation{Center for Astrophysics $|$ Harvard \& Smithsonian, 60 Garden Street, Cambridge, MA 02138, USA}
\email[]{gfazio@cfa.harvard.edu}

\author[0000-0001-8921-3624]{Nicole M. Ford}
\affiliation{McGill University, Montreal QC H3A 0G4, Canada}% 
\affiliation{Trottier Space Institute, 3550 Rue University, Montréal, Québec, H3A 2A7, Canada}
\email[]{nicole.ford@mail.mcgill.ca}

\author[0000-0001-6803-2138]{Daryl Haggard}
\affiliation{McGill University, Montreal QC H3A 0G4, Canada}% 
\affiliation{Trottier Space Institute, 3550 Rue University, Montréal, Québec, H3A 2A7, Canada}
\email{daryl.haggard@mcgill.ca}

\author[0000-0002-5599-4650]{Joseph L. Hora}
\affiliation{Center for Astrophysics $|$ Harvard \& Smithsonian, 60 Garden Street, Cambridge, MA 02138, USA}
\email{jhora@cfa.harvard.edu}

\author[]{Howard A. Smith}
\affiliation{Center for Astrophysics $|$ Harvard \& Smithsonian, 60 Garden Street, Cambridge, MA 02138, USA}
\email{hsmith@cfa.harvard.edu}

\author[0009-0004-8539-3516]{Zach Sumners}
\affiliation{McGill University, Montreal QC H3A 0G4, Canada}% 
\affiliation{Trottier Space Institute, 3550 Rue University, Montréal, Québec, H3A 2A7, Canada}
\email[]{ronald.sumners@mail.mcgill.ca}

\author[0000-0003-2618-797X]{Gunther Witzel}
\affiliation{Max Planck Institute for Radio Astronomy, Bonn \& 53121, Germany}
\email{gwitzel@mpifr-bonn.mpg.de}

\begin{abstract}
We determine the mid-infrared (MIR, $\sim$5~\micron--22~\micron) extinction towards the Galactic center
using MIRI/MRS integral field unit (IFU) observations of the central $3''\times3''$ region (near 5~\micron) to $7''\times7''$ region (near 22~\micron).
To measure the MIR extinction, we employ two approaches: modeling the intrinsic-to-observed dust thermal spectrum and assessing the differential extinction between hydrogen recombination lines.
Expanding on prior work, we directly model the dust-opacity distribution along the line of sight, and we make available a Python code that provides a flexible tool for deriving intrinsic dust emission spectra.
We confirm the spatial variability of extinction across the field, demonstrating that dusty sources---such as IRS~29N---exhibit higher local extinction.
Furthermore, we verify the absence of PAH emission features in the Galactic center MIR spectra.
Using the two complementary methods, we derive a refined “best guess” MIR extinction law for Sgr~A* and the surrounding Galactic-center region. By applying the extinction law to a MIR flare measurement discussed in a companion paper \citep{Michail_SED}, we estimate a residual relative extinction uncertainty for the short MIRI/MRS grating on the order of $0.2~\mathrm{mag}$ {from $\sim$5 to $\sim$18~\micron\ and $\sim$0.3~mag from $\sim$18 to $\sim$22~\micron},  consistent with our uncertainty estimate.
\end{abstract}

\keywords{\uat{Galaxies}{573} --- \uat{High Energy astrophysics}{739} --- \uat{Interstellar medium}{847}}


\section{Introduction} 
The Galactic center has long been the subject of intense study owing to its unique suitability to probe fundamental astrophysical processes such as accretion on to supermassive black holes (SMBHs), stellar dynamics, and tests of fundamental physics. All X-ray, ultraviolet (UV), visible, and infrared astronomical observations of the Galactic center are inherently limited by the extinction ($A_\lambda$) caused by dust grains along the line of sight. 
In the UV and visible ($\lambda <  1~\mathrm{\mu m}$), the extinction prohibits observations altogether \citep[$A_\lambda \sim 30$; e.g.,][]{Becklin1968,Becklin1978,Rieke1985}. At infrared wavelengths, the extinction declines quickly and reaches a local minimum somewhere between 4 and $8~\mathrm{\mu m}$ \citep[e.g.,][]{Lutz1996}. This leads to the $K$-band (2.0--2.4~\micron) being the workhorse spectral window for ground-based observations because of its modest (for the Galactic center) extinction of $\sim$2.2--2.5~mag \citep{Schodel2010, Fritz2011} and much lower thermal background than at longer wavelengths. 
Modern $K$-band instruments \citep[e.g.,][]{Lenzen2003, Eisenhauer2003, Larkin2006, Wizinowich2006, GRAVITY2017} have allowed {precision}  measurements of stellar orbits in the vicinity of the SMBH \citep{Ghez2008, Gillessen2017, Do2019, GRAVITYCollaboration_schwarzschild} and detection of a flare orbiting near the innermost stable circular orbit \citep{GRAVITYCollaboration2018}.

Despite extensive study, the exact shape of the Galactic center extinction curve is still uncertain.
Water ices increase the extinction by roughly one magnitude around $3.05~\mathrm{\mu m}$, after which it drops and reaches its lowest value ($A_\lambda \sim 1~\mathrm{mag}$) between $4~\mathrm{\mu m}$ and $8~\mathrm{\mu m}$. However, molecular resonances of $\mathrm{CO}$, $\mathrm{CO_2}$, $\mathrm{H_2O}$ lead to narrow spikes in the extinctions \citep{Lutz1996, Fritz2011} in this wavelength range.
Starting at $\sim8~\mathrm{\mu m}$, silicate absorption leads to a strong increase of extinction which reaches a local maximum $A_\lambda \sim  4$--5 at $9.8~\mathrm{\mu m}$. Finally, a second silicate absorption feature at ${\sim} 18~\mathrm{\mu m}$ increases the extinction again, beyond which it drops to near $0$ (following a power law $\sim \lambda^{-2}$; \citealt{Draine2001}).

Extinction poses two challenges to infrared observations of the Galactic center: the extinction removes photons and thus decreases sensitivity, especially for observations at  $\lambda \la2~\mathrm{\mu m} $ and in the 8--13~\micron\ silicate absorption band. Perhaps more important, uncertainty in the extinction limits photometric accuracy because the extinction uncertainty is much larger than the measurement uncertainty for all but the faintest sources \citep[e.g.,][]{Schodel2010, Fritz2011}.
The uncertain extinction, and especially its uncertain wavelength dependence, is particularly important for studies of the spectral energy distribution (SED) of \Sg\ \citep[e.g.,][]{Witzel2018,VonFellenberg2018,GravityCollaboration2020flux,Michail2024}. This uncertainty is much larger than the measurement uncertainty and remains the limiting factor even as ever-increasing instrumental capabilities \citep[e.g.,][]{GravityCollaboration2021_xrayflare,Schoedel2013,Weldon2023} allow novel measurements of the source. 
%A related aspect is the spectral index measurements of the source for which the extinction poses the prime observational uncertainty \citep[e.g.,][]{Michail2024}.

Three prime methods have been used to estimate the Galactic center extinction:
\begin{enumerate}
    \item Stellar colors \citep{Becklin1968,Nishiyama2006,Nishiyama2008,Nishiyama2009,Schodel2010,Nogueras-Lara2018,Nogueras-Lara2020};
    \item IR-line ratios compared to radio free-free emission \citep{Willner1983,Lutz1996,Scoville2003,Fritz2011}; and
    \item IR-line ratios compared to radio recombination lines \citep{Scoville2003}.
\end{enumerate}
To measure extinction in sources other than the Galactic center, and in particular for extragalactic sources, models fit the observed SED with a derived intrinsic emission modified by a chosen extinction curve. Popular codes to compute such continuum SED models include PAHFIT \citep{Smith2007pahfit} and CAFE \citep{Marshall2007}. For the intrinsic emission, these codes typically use a mix of dust continuum emission at different temperatures and sets of polycyclic aromatic hydrocarbon (PAH) emission  profiles. The latest high S/N and spectral resolution JWST/MIRI spectra of such sources have shown the limitations of these approaches \citep[e.g.,][]{Donnan2023}, and mitigation strategies are being developed \citep[e.g.,][]{Donnan2024}.

While extinction estimates in the near-infrared, in particular the $K$-band, derived by \cite{Fritz2011} and \cite{Schodel2010} serve as the {\it de facto} standard of the community, no high-spatial-resolution extinction values for the mid-infrared exist. The latest work measuring mid-infrared extinction values for the Galactic center is that by \cite{Fritz2011}, who used ISO/SWS data originally published by \cite{Lutz1996}. These authors used hydrogen recombination-line ratios and the continuum spectrum observed by ISO/SWS with a beam size of roughly 14\arcsec$\times$16\arcsec. 
However, spatially resolved extinction measurements by \cite{Nishiyama2006,Nishiyama2008,Nishiyama2009,Schodel2010} and \cite{Scoville2003} indicate that extinction can vary  spatially by up to 1~magnitude at $K$\null. An analysis of $M$-band spectra of several IRS sources reveal local absorption features which could be attributed to the stellar envelopes \citep{Moultaka2005,Moultaka2019}.
A recent analysis by \cite{Haggard2024} even demonstrated that the extinction may be \textit{temporally} variable on $\sim$year timescales as local absorbing features orbit through our line of sight.

In order to interpret modern observations, a high-fidelity extinction correction through the MIR at the spatial resolution of JWST is required. This paper is a first attempt to derive the mid-infrared extinction law to the Galactic center region from MIRI/MRS spectra. 
This work combines continuum modeling (Section~\ref{sec:continuum_modeling}) with line-derived extinction measurements (Section~\ref{sec:line_modeling})  to estimate both the spatial variations and an extinction law for the line of sight to \Sg. Section~\ref{s:obs} describes the observations used, and Section~\ref{s:sum} summarizes the current status and future prospects.

\section{Observations, Data, and Calibration}
\label{s:obs}
\subsection{JWST MIRI/MRS Data}
The MIRI Medium-Resolution Spectrometer\footnote{\url{https://jwst-docs.stsci.edu/jwst-mid-infrared-instrument/miri-observing-modes/miri-medium-resolution-spectroscopy}} \citep[MRS;][]{Wells2015} observations used the Integral Field Unit (IFU) to obtain complete spectra in the 4.9 to 27.9~\micron\ range.  The MRS has four spectral channels that observe simultaneously, but each channel observes only 1/3 of its available spectral range at a time. Obtaining a complete spectrum requires observations at three different grating angles (designated short, medium, and long). The IFU's field of view (FoV) is 3\farcs2$\times$3\farcs7 in channel~1 (wavelengths $\lambda<7.65$~\micron) and increases in steps for each channel to  6\farcs6$\times$7\farcs7 for channel~4 ($\lambda>17.9$~\micron). Pixel sizes range from 0\farcs196 to 0\farcs273, and spectral resolution ranges from $\sim$3500 at short wavelengths to $\sim$1500 at long wavelengths. The observations used here \citep[PID 4572,][]{2023jwst.prop.4572H} were all centered on \Sg\ (J2000 coordinates $17^{\rm{h}}45^{\rm{m}}40\fs04,-29^\circ00'28\farcs17$) and were taken in 2024 Apr 6--9 and Sep~6. Of the four planned full-spectrum observations (channels 1--4 with the short, medium, and long grating angles), only two were completed during this program's execution, giving 333~s of exposure time in each wavelength channel at each grating setting. There were also eleven staring (undithered) observations to monitor the MIR variability of \Sg. These used channels 1--4 and provided a total exposure time of 22.3 hours per channel but observed only the short grating  (wavelength bands 4.90--5.74, 7.51--8.77, 11.55--13.47, and 17.70--20.95~\micron).
    
Data calibration began with the Level~1 calibrated RATE files, which manually passed through the Level~2 pipeline
version 1.17 \citep{2023zndo...6984365B} in context \texttt{pmap} version 1322. This applied the two-dimensional residual fringing correction,\footnote{\url{https://jwst-pipeline.readthedocs.io/en/latest/jwst/residual_fringe/main.html}} critical to mitigating the detector-based fringing of the spectra. Subsequent Level~3 calibration revealed differences in the locations of the point sources in the field ranging from $<$1 pixel to several channel~1 pixels. To compute the relative shifts, we summed the Level~3, channel~1 spectral data from each observation and chose the ``O11'' full-dithered observations as the spatial reference. Using full-dithered rather than a staring observation mitigated the effects of the undersampled PSF in undithered observations. Cross-correlating each of the other channel~1 observations with the reference produced a pointing ASDM offset file for each. We then reprocessed the Level~2 files through the Level~3 pipeline including the  offsets to produce a data cube for each observation and each channel with that channel's default spaxel resolution (0\farcs13, 0\farcs17, 0\farcs20, and 0\farcs35 for channels 1--4, respectively).  These modestly sub-sample the physical pixels.

To combine the separate observations, several post-pipeline steps were needed. The first was to convolve each observation to a uniform spatial resolution using \texttt{astropy.convolve} \citep{astropy:2022}.  The convolution kernel was $\rm FWHM = \sqrt{FWHM_T^2 - {FWHM}_\lambda^2}$, where FWHM$_{\rm T}$ is the desired FWHM, and FWHM$_\lambda$ is  the FWHM of the spectral slice being convolved. FWHM$_\lambda$ came from \cite{Law2023}, who assumed a two-dimensional Gaussian PSF\null.

\begin{figure}
    \centering
    \includegraphics[width=0.485\textwidth,trim=0cm 0cm 1cm 0cm, clip]{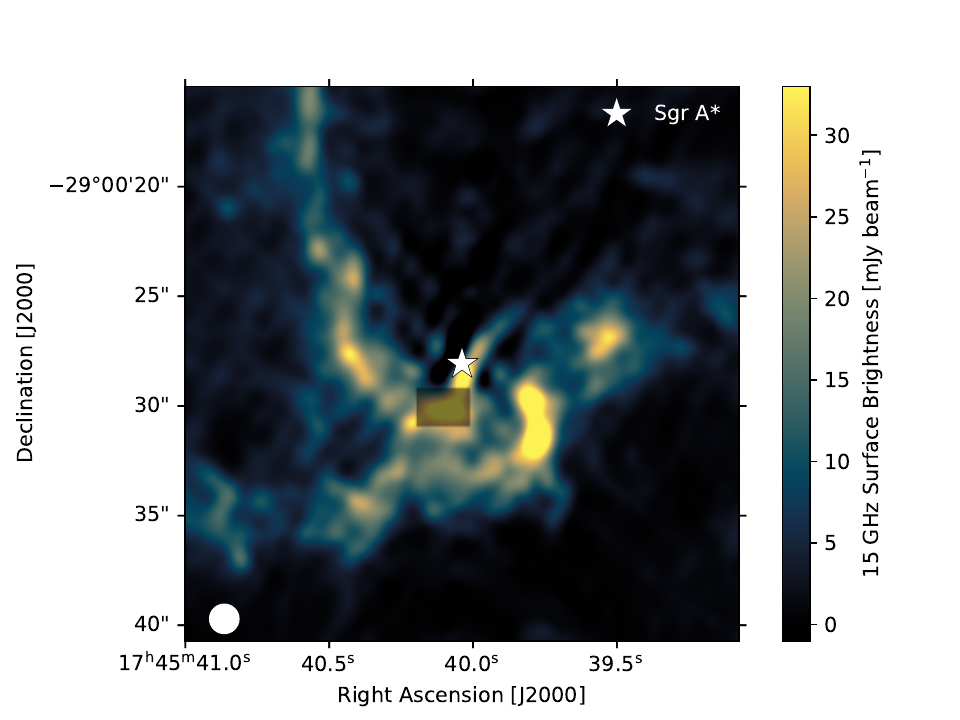}
    \caption{15~GHz VLA image. The grey box indicates the region used to model the \Ms\ extinction, and the location of \Sg\ is indicated with a white star. The white circle in the bottom left indicates the beam size smoothed to JWST/MIRI 17.98 $\mu$m resolution.}
    \label{fig:minispiral_extinction}
\end{figure}

The second step was to calibrate each observation in flux density. This required identifying a region of high but constant surface brightness, and for that we chose a $3\times3$ spaxel box in the \Ms\ centered on 17$^{\rm{h}}$45$^{\rm{m}}$39\fs93, $-$29\degr00\arcmin28\farcs54 (Figure~\ref{fig:minispiral_extinction}). Because the full-dither and stare observations have differing spectral coverage, we matched the average flux in short grating wavelength range for each of the four channels. We chose the ``O11'' observation, the same used above for the relative offset calculations, as our flux reference. The multiplicative scale factors were 0.99--1.17 in channel~1, 1.00--1.09 in channel~2, and 0.98--1.02 in channels~3 and~4.

The third step was to re-project each observation to the same spatial WCS as the channel~4 grid using the flux-conserving \texttt{reproject\_exact} function in the \texttt{astropy} \texttt{reproject} package. This gave spatial pixel sizes of 0\farcs35. 

The final reduction step was to spatially average the WCS-matched cubes while maintaining the spectral resolution produced by the pipeline. For the Section~\ref{sec:continuum_modeling} continuum analysis, which used \textit{only} the two full-spectrum, four-point-dither observations, this meant convolving channels~1--4 to  0\farcs87 to match the resolution at 23~\micron. For the Section~\ref{sec:line_modeling} line analysis, for which higher spatial resolution is more beneficial, channels~1--3 were convolved to 0\farcs70, the resolution  at 17.98~\micron,  but channel~4 retained its native 0\farcs73 resolution (corresponding to the longest-wavelength line at 19.06~\micron).

\subsection{VLA: 15 GHz Free-Free Emission}
\cite{Scoville2003} and \cite{Fritz2011} utilized 15~GHz emission, which is dominated by free-free emission, to estimate the intrinsic fluxes of hydrogen recombination lines. The ratio of observed line fluxes to the intrinsic ones then gives the line's absolute extinction towards the Galactic center. Both earlier analyses used data from the pre-upgrade Karl G. Jansky Very Large Array (VLA), which had spectral bandwidth limited by the original backend correlator.

Newer 15 GHz VLA observations are now available.  The upgraded correlator has wider bandwidth and therefore better sensitivity. \Sg\ was observed in the A~configuration (PID VLA/14A-231) on 2014 March 10 yielding 4.49~hours of on-source time. These data were taken in 8-bit correlation with 16 spectral windows, each with 64 2-MHz-wide channels, giving a total bandwidth of 2048~MHz. We processed these data through the standard VLA pipeline in CASA \citep{McMullin2007} using J1331+3030 as the flux calibrator, J1733$-$1304 as the bandpass calibrator, and J1744$-$3116 as the complex gain calibrator. The calibrated data underwent two rounds of phase self-calibration, first with a point-source model, then self calibrating based on CLEAN-ed visibilities. 
The C-configuration data (PID VLA/16A-274) were obtained on 2016 April 21 also phase-centered on \Sg\ for an on-source time of 5.15~hours. These data are in 3-bit mode consisting of 64 spectral windows with 128 channels, each with 0.5-MHz width, for a total bandwidth of 4096~MHz. We processed these data through CASA using the default VLA pipeline with J1331+3030 as the flux and bandpass calibrator,  J1744$-$3116 as the complex gain calibrator, and the same phase self-calibration steps as for A-configuration. 

\Sg's minute-timescale variability and overall brightness relative to the \Ms\ create artifacts in radio images of the Galactic center.  These artifacts make absolute-flux measurements of the surrounding regions, where the emission lines are seen, uncertain. To ameliorate this, we subtracted the average flux of \Sg\ during each observation using \texttt{uvsub} in CASA\null. The flux and spectral index of \Sg\ were obtained from an ongoing VLA study of its spectrum and variability over the last decade (J. Michail et al., in prep.). In short, each spectral window is imaged per observation, and \Sg's flux is calculated by fitting a point source to the phase center. These data are then fit with a power-law ($F_\nu\propto\nu^{\alpha}$) to determine the flux at standard frequency and the spectral index across the observing band. On 2014 March 10, \Sg's 15~GHz flux density was $F_{15}= 1.03$~Jy with $\alpha=0.33$, and on 2016 April 21, it was $F_{15} = 0.92$~Jy with $\alpha=0.34$; these fluxes and spectral indices are used to model and subtract \Sg\ from their respective observations. The \Sg-subtracted MeasurementSets were then concatenated with \texttt{concat}, and \texttt{statwt} was used to re-normalize the statistical imaging weights between the datasets. The concatenated dataset yields baselines ranging from 14.27~m to 36.62~km (0.7--1832~k$\lambda$ at 15~GHz).

We created the 15~GHz image with \texttt{TCLEAN} using the multi-term, multi-frequency synthesis (MTMFS) deconvolver with $\textsc{nterms}=2$. The data were Briggs-weighted with $\textsc{robust}= 0.5$ to give a restoring beam of 0\farcs29$\times$0\farcs15 (position angle 7\fdg7). The final image has a pixel scale of 0\farcs04 and image size 5832$\times$5832~pix$^2$ $\approx$ $233\arcsec\times233\arcsec$. CLEANing was non-interactive and stopped when the noise in the residual image reached $4\sigma$. Because of the large extended emission from the \Ms, we used \texttt{scales=[0, 5, 20]} pixels (physical sizes point source, 0\farcs2, and 0\farcs8) to speed up convergence to a final image. The final RMS noise  $\sigma = 19.3$~$\mu$Jy beam$^{-1}$ before primary-beam correction with \texttt{widebandpbcor}.  This task was required because of the large frequency range. Finally, we Gaussian-smoothed the native-resolution VLA image with \texttt{imsmooth} to 0\farcs57 to match MIRI/MRS's FWHM at 17.98~$\mu$m for the  Section~\ref{sec:line_modeling} line analysis.  Figure~\ref{fig:minispiral_extinction} shows the image.
    
% \subsection{ALMA: H42$\alpha$ Recombination Line}
%     We obtained continuum-subtracted spectra of the H42$\alpha$ line (rest frequency 85.6883 GHz\footnote{Obtained from \href{https://splatalogue.online/\#/home}{Splatalogue}}) from the ALMA archive from project 2023.1.00295.S (PI: M. Araki) observed on 5 and 23 December 2023 with the 12-meter array. These data have a total time-on-source of 86.3 minutes. The data were processed through the default ALMA pipeline (version 2023.1.0.124) in CASA 6.5.4.9, self-calibrated, and the continuum baseline was manually determined during QA2. The spectral window (SPW) containing the H42$\alpha$ emission is centered at a sky frequency of 86.003 GHz with a total bandwidth of 1.875 GHz and 3840 spectral channels (0.49 MHz spectral resolution; $1.71$ km s$^{-1}$ velocity resolution at 85.6883 GHz). The primary-beam corrected cube has pixel scale 0\farcs096, image size 1200$\times$1200 pixels (115\farcs2), a final restoring beam 0\farcs756$\times$0\farcs476 (76.8\degr position angle), and contains H42$\alpha$ velocities between -4366 to 2185 km s$^{-1}$. The (non-primary-beam corrected) RMS per spectral plane is approximately 0.75 mJy beam$^{-1}$.
    
\section{Continuum Modeling}\label{sec:continuum_modeling}
Continuum modeling provides one way to estimate the line-of-sight extinction. Models are based on integrating the equation of radiative transfer along each line of sight of interest. Model parameters for galaxies typically include
a dust temperature distribution, 
a dust emissivity model, 
and an extinction description.
The dust emissivity may include PAH emission, and a stellar contribution (essentially Rayleigh--Jeans for the wavelengths considered here) can be added to the continuum model. 
The observations reported here show no PAH emission in the spectra, consistent with previous MIR observations of the Galactic center \citep[e.g.,][]{Lutz1996,Simpson2007}. This is likely related to the high UV radiation field in the region \citep[e.g.,][]{Lutz1996,Paumard2006,Simpson2007}.
In addition, stellar emission is marginal for the wavelengths considered here, and stars are typically detected only if they illuminate dusty envelopes (as in the cases of IRS~16C, IRS~16NW, and IRS~29N). 

In order for models to converge, at least some of the parameters have to be fixed or at least constrained based on other information. In the models used here, the fixed parameter is the emissivity of the emitting dust
$\epsilon(\lambda)$. Dust emissivity is approximately a power law, typically $\epsilon(\lambda)\propto\lambda^{\beta}$ with $\beta$ between $-1.5$ and $-2$ \cite[e.g.,][]{Smith2007pahfit}. However, in the mid-infrared, light absorbed at shorter wavelengths is re-emitted at longer wavelengths, causing an increase of the effective emissivity \cite[e.g.,][]{Li2001,Draine2001,Draine2007}. The strength of this reprocessed emission depends on dust temperature $T$ \citep[e.g.,][]{Marshall2007}, which means that $\epsilon(\lambda)$ is degenerate with the {\it a priori} unknown temperature distribution $\Psi$ of the dust.  Towards the Galactic center, $T\approx200~\mathrm{K}$ \citep[e.g.,][]{Dinh2024}, for which the dust emissivity is not negligible \citep{Marshall2007}. 
Because of that, following \cite{Donnan2024}, we fixed $\epsilon(\lambda)$ to the values reported by \cite{Li2001} (and which we copied from \citealt{Donnan2023}). 
The derived extinction values will depend on this choice. Specifically, this law has a $\epsilon(\lambda \to 0) \propto \lambda^{-1.5}$ asymptotic behavior.

With the dust emissivity fixed, the dust emission at any location along the line of sight is $\epsilon(\lambda)B_\nu(T,\lambda)$, {where $B_\nu(T,\lambda)$ is the Planck function}. This emission is extincted by wavelength-dependent optical depth $\tau(\lambda)$ of absorbing and scattering dust along the line of sight \citep[e.g.,][]{Donnan2024}.  The continuum flux density $C(\lambda)$ is given by
\begin{equation}
\begin{split}
    C(\lambda) =&
    A \int_0^\infty \int_{35~\mathrm{K}}^{1500~\mathrm{K}} \Psi (T, \tau'_{9.8}) \dfrac{\epsilon(\lambda) B_{\nu}(T, \lambda)}{B_0(T)}\\ & e^{-\tau'_{9.8} \tau_\lambda} dT d\tau'_{9.8}\quad,
\end{split}
\label{eq:dust_continuum}
\end{equation}
where {the integration variable} $\tau'_{9.8}$ represents optical depth along the line of sight at 9.8~\micron, $\tau_\lambda$ is the wavelength dependence to be solved for, $\tau(\lambda)=\tau'_{9.8}\tau_\lambda$ is the optical depth along the line of sight at wavelength $\lambda$, and $T$ is an integration variable representing the dust-temperature distribution at each $\tau'_{9.8}$.  
The lower integration temperature bound is based on the requirement to contribute MIR flux, and the upper bound corresponds to the typical dust sublimation temperature estimated for AGN \citep{Kishimoto2007, Donnan2024}. Following \cite{Donnan2024}, we set the temperature normalization $B_0(T) = \int_{5\mu m}^{28\mu m} \epsilon(\lambda) B_{\nu}(T, \lambda) d\lambda$. 
The function $\Psi(T, \tau'_{9.8})$ is the two dimensional distribution function that describes the dust temperature distribution and the absolute extinction distribution. 

Codes such as PAHFit \citep{Smith2007pahfit} choose $\Psi(T,\tau'_{9.8})$ to be series of a single-temperature  blackbodies, where typically up to $10$ temperature components are chosen. 
This approach worked well for lower-fidelity spectra obtained with Spitzer IRAC, but \cite{Donnan2023,Donnan2024} showed that the approach is insufficient for high-sensitivity, high-spectral-resolution MIRI/MRS data. 
Instead, they inferred the distribution $\Psi
(T,\tau'_{9.8})$ directly from the data by fitting a $20\times20$ grid and imposing continuity of $\Psi$ through regularization of the optimization cost function. 

The approach used here is to force $\Psi(T, \tau'_{9.8})$ to be a two-dimensional Gaussian field (GF). This is similar to the \cite{Donnan2024} method but has the advantage that  standard methods can be used for deriving the GF power spectrum, and no likelihood regularization has to be enforced. 
This allows more degrees of freedom (DoF) in the fit, and a logarithmically-spaced ${\rm GF}(T\times\tau)$ field works well (dimensions $80\times50=4000~\mathrm{DoF}$). 
%
%The extinction was modeled as $e^{-\tau_{9.8}\tau_\lambda}$ in \autoref{eq:dust_continuum}. $\tau(\lambda)$ is the normalized optical depth profile of the dust, scaled to the optical depth at $\tau(\lambda=9.8)$. 
\cite{Donnan2023} derived an extinction curve $\tau_\lambda$ based on JWST observations of dust-obscured galaxies, but their $\tau_\lambda$ shows substantial differences from other such profiles \citep[e.g.,][]{Kemper2004, Ossenkopf1992, Fritz2011,Jones2013}. 
Because of this, we treated the optical depth as a free parameter, which we inferred from the data by allowing for a deviation ($\delta \tau_\lambda$) from the \cite{Kemper2004} curve. 
Explicitly, we modeled $\delta \tau(\lambda)$ as an additional one-dimensional GF, $\tau_\lambda = \tau_{\lambda,\rm{Kemper2004}} \delta \tau_\lambda$. 

For the actual fitting, we flagged bright emission lines and binned the flagged, observed spectra into $75$ wavelength bins. This greatly reduced the numerical complexity. We transformed the data into logarithmic flux units and added a $3\%$ normally distributed scatter to the data. This added uncertainty is much larger than the measurement error, which is dominated by residual systematic errors from the fringing. Adding a well-known uncertainty to the data improved numerical performance and avoided biases due to fringing and other data artifacts.
The implementation of the Gaussian fields was done with the \textsc{nifty8.re} framework \citep{Arras2021,Arras2022,niftyre}, using the optimization methods detailed by \cite{Frank2021} and \cite{knollmüller2020metric}. This approach allows  generating posterior samples, which can be used to assess the  uncertainty of the derived extinction. A notebook to derive the extinction for MIRI/MRS spectra is available on GitHub.\footnote{\url{https://github.com/Sebastiano-von-Fellenberg/MIR-Extinction/tree/main}}

\begin{figure*}
    \centering
    \includegraphics[width=0.985\linewidth]{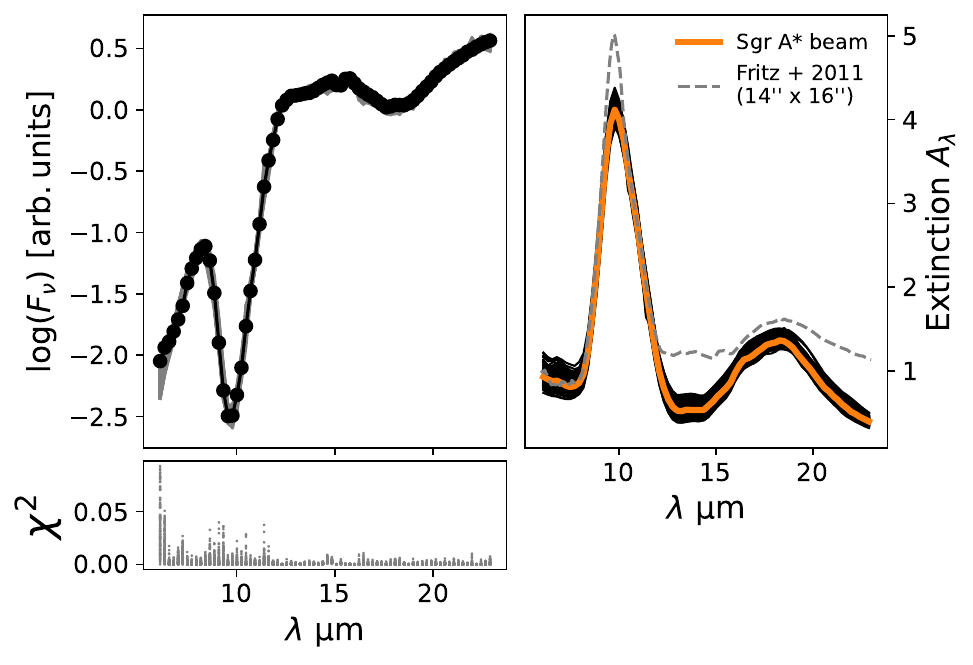}
    \caption{Continuum model (top left) and derived extinction law (top right) derived from {a fit to} the spectrum in the \Sg\ pixel. The black line and dots show the binned spectrum with an artificially added $3\%$ log-normally distributed {uncertainty}. The grey lines show posterior samples of the extincted dust model. The panel below shows the $\chi^2$ residual. The right plot shows the derived extinction curves ($A_\lambda$). The orange line shows the median posterior model, and the black line shows the posterior samples. The dashed-grey line shows the extinction law derived by \cite{Fritz2011}. That law is based on the ISO/SWS observations by \cite{Lutz1996}, which had an effective beam size of $14''\times16''$.}
    \label{fig:extinction}
\end{figure*}

\begin{figure*}
    \centering
    \includegraphics[width=0.85\textwidth]{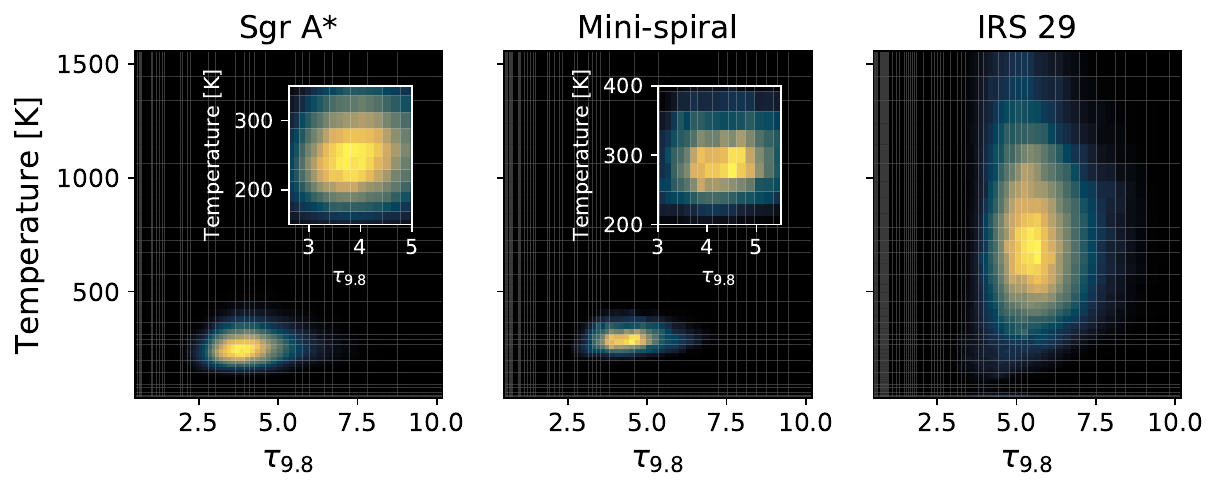} %for virdis remove _eclipse
    \caption{Dust distribution functions $\Psi(T, \tau_{9.8})$ for the \Sg\ pixel (left), the \Ms\ (center), and IRS~29 (right). The insets in the \Sg\ and \Ms\ panels show zoom-ins on the peaks of the respective posterior distributions.}
    \label{fig:dust_properties}
\end{figure*}

\subsection{\Sg\ fit and results}
For \Sg, the spectrum to be fit was the flux density in the single pixel at \Sg's position. The fitting used $30\times4=120$ random seeds, some of which led to obvious poorly converged fits.  The continuum model in  \autoref{fig:extinction} is based on the 50\% of posterior samples showing the least $\chi^2$. The largest residuals are at $\lambda < 6~\mathrm{\mu m}$.  These are plausibly explained by the contribution of stellar emission, but adding a power-law stellar-emission component gave residuals much larger than the model uncertainty. A plausible explanation is that the data are insufficient to fit the dust and stellar emission simultaneously. 
As a workaround, we excluded data shorter than $6~\mathrm{\mu m}$. Future work should include NIRSpec data to simultaneously model the near-infrared and MIR continuum. 

The dust distribution $\Psi(T, \tau_{9.8})$ along \Sg's line of sight is well approximated by a narrow temperature range around $T\approx 250~\mathrm{K}$ as shown in Figure~\ref{fig:dust_properties}. This is consistent with  previous studies \citep[e.g.,][]{Schodel2011,Dinh2024}. 
The width of the temperature distribution is poorly constrained with both very peaked functions of $T$ and broader ones fitting the data well. 
The opacity distribution is centered on $\tau_{9.8}\sim3.8$, but the uncertainty is about 0.5.

\subsection{\Ms\ fit and results}\label{sec:conti_mini_irs29}
The \Ms\ input spectrum was the sum of flux densities {within} the  aperture shown in \autoref{fig:minispiral_extinction}. 
The \Ms\ fits are less constrained than the ones for \Sg\ as shown in Figures~\ref{fig:dust_properties} and~\ref{fig:dust_cont} but are consistent with the findings of previous studies \citep[e.g.,][]{Schodel2011,Dinh2024}.  The dust opacity $\tau_{9.8}\sim4.6$, slightly higher than for \Sg, but the difference is barely significant. The extinction curve $A_\lambda$ is similar to that of \Sg, but there may be slightly more extinction shortward of the $\lambda=9.8~\mathrm{\mu m}$ silicate absorption feature. 

\subsection{IRS~29 fit and results}
\begin{figure}
    \centering
    \includegraphics[width=0.5\textwidth]{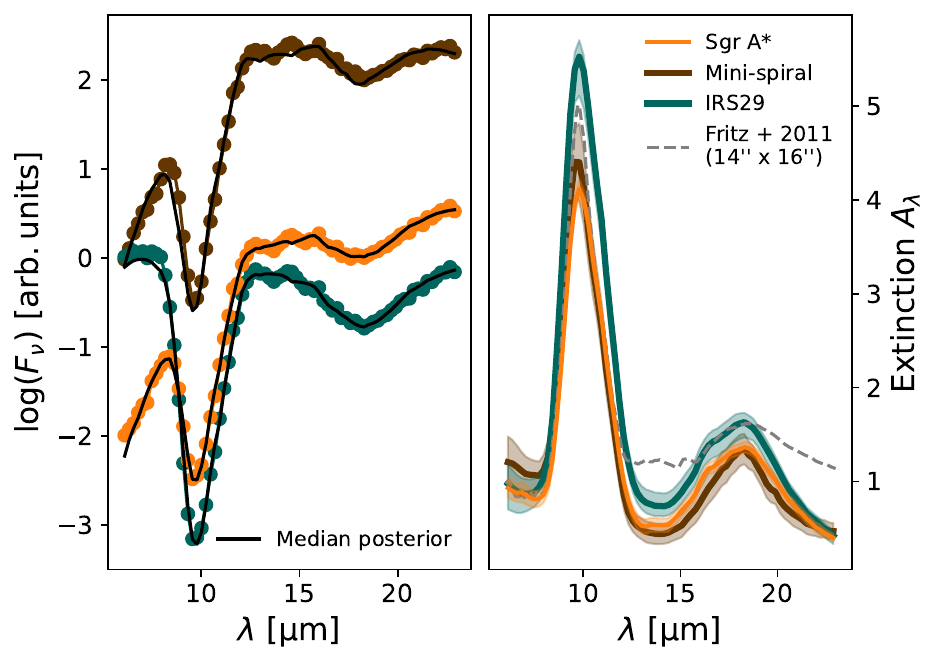}
    \caption{Dust continuum models and derived extinction laws for different regions observed with JWST MIRI/MRS\null. The left panel shows the \textbf{continuum} fits to apertures centered on \Sg, the Mini-spiral, and the MIR-bright star IRS~29N. The right panel shows the corresponding extinction laws. The grey-dashed line shows the MIR-extinction law derived by \citet{Fritz2011}.}
    \label{fig:dust_cont}
\end{figure}

The IRS~29 input spectrum was the sum of a 3$\times$3-pixel region centered on its brightest pixel.
The extinction values are higher than for the other two sources, $\tau_{9.8}\approx5.5$ (Figures~\ref{fig:dust_properties} and~\ref{fig:dust_cont}). Higher extinction is seen across the entire wavelength range with the possible exception of shortward of the $\lambda=9.8~\mathrm{\mu m}$ silicate absorption feature.  The dust temperature distribution is significantly warmer, $T\approx800$--1000~K, and broader than for the other two sources, consistent with thermal dust emission from a stellar envelope at a range of radii. Evidence for local absorption was found in $M$-band spectra of the star \citep{Moultaka2019}. The temperature distribution extends to $T>1000$~K, and a stellar photosphere contribution at wavelengths $\la6.5~\mathrm{
\mu m
}$ cannot be ruled out. 

\subsection{Dust Opacity}
%We show the median posterior distribution of the function $\Psi(T, \tau_{9.8})$ for \Sg, the \Ms\ and IRS~29 in \autoref{fig:dust_properties}.

%For both the \Ms\ and the \Sg,  the function $\Psi(T, \tau_{9.8})$ are well approximated by a single temperature dust distribution with $T_{\rm{dust}}\approx 200 - 300~\mathrm{K}$, which is consistent with the finding of previous studies \citep[e.g.,][]{Schodel2011,Dinh2024}. The mini-spiral has slightly higher temperature $T\sim 300~\mathrm{K}$, while the \Sg\ pixel is colder with $T\sim 250~\mathrm{K}$. 
%Draws from the posterior sample show that the range of allowed distributions is relatively broad, with both very peaked functions of $\Psi(T,\tau_{9.8})$ as well as broader distributions fitting the data well. 

%For both the the Mini-spiral and Sgr~A*, the opacity distributions are centered on $\tau_{9.8}\sim4$, with slightly higher values for the \Ms\ which is reproduced in the effective extinction curve (see right panel of \autoref{fig:minispiral_extinction}). 
%The dust distribution of IRS 29 is different, with higher temperatures $\sim800-1000~\mathrm{K}$, and a higher opacity of $\tau_{9.8}\sim 5$. 

\begin{figure}
    \centering
    \includegraphics[width=0.485\textwidth]{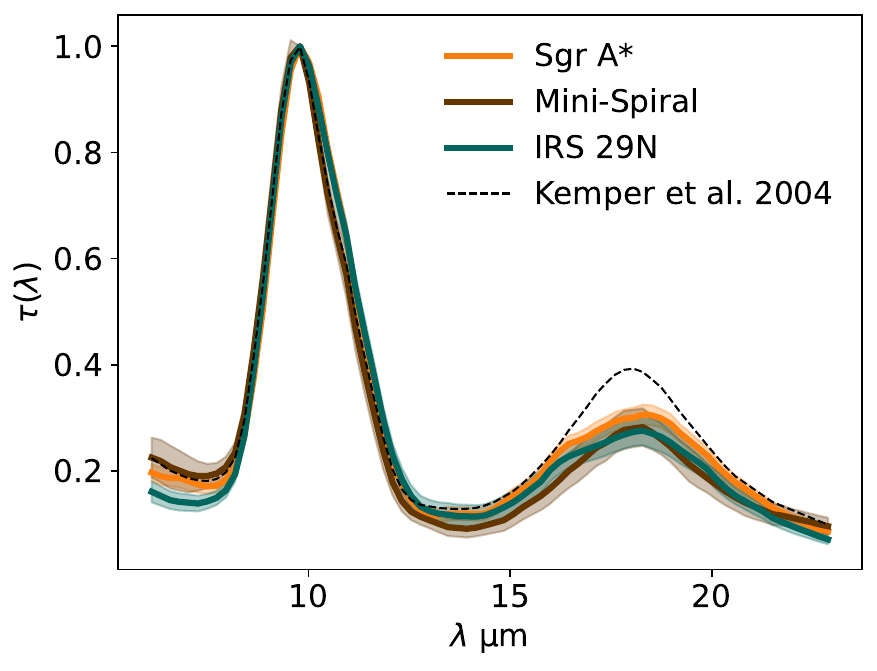} 
    \caption{Posterior distribution of the optical depth law $\tau(\lambda)$ defined in \autoref{eq:dust_continuum}. Different colors show the optical depths for different regions. The dashed line shows the optical depth profile by \cite{Kemper2004}, which serves as a prior model and from which we allowed a multiplicative deviation.}
    \label{fig:opacity}
\end{figure}

%\subsection{Results -- Optical depth}
%As discussed in \autoref{sec:continuum_modeling}, we allow for a multiplicative deviation $\delta \tau(\lambda)$ from the prior optical depth profile ($\tau_{\rm{Donnan2023}}$, \citealt{Donnan2023}).
\autoref{fig:opacity} shows the posterior  optical depth profile $\tau_\lambda$ for \Sg, the \Ms, and IRS~29. The three profiles agree well overall, and they agree with the \cite{Kemper2004} prior model at most wavelengths. However, there is a significant deviation from the prior at $\sim$19~\micron, i.e., the longer-wavelength silicate  absorption feature. \cite{Donnan2024} found similar variability between the ratio of the $9.8~\mathrm{\mu m}$ and $18~\mathrm{\mu m}$ absorption features for different AGN sources. They argued that this variability may be related to differing relative obscuration and temperature differences in the respective dust distributions. 

% yes, I will add a discussion here in a second. just at the end of section 5.5. in Donnan 2024 they discuss it. i think we can write the same...

%% hi steven, the optical depth prior model that we use is definetly from Kemper 2004, not Donnan 2023. I will correct the  legend figure soon, please keep an eye out if we still have donnan 2023... 

\section{Line-Derived Extinction}\label{sec:line_modeling}
\subsection{Line Measurements}
A wholly independent measure of extinction comes from the observed hydrogen recombination lines.
%MIRI/MRS offers high spatial resolution and high sensitivity IFU observations which make it ideally suited to repeat the line-derived extinction measurements obtained for 
Based on standard approaches \citep[e.g.,][]{Osterbrock1989}, \cite{Scoville2003} and \cite{Fritz2011} applied this method to the Paschen-$\alpha$ and Brackett-$\gamma$ lines, and MIRI/MRS extends it into the MIR\null. 
The MIR line ratios give relative extinction measurements between lines, and the ratio between any line flux and the radio-continuum emission at the same location gives 
an absolute extinction at that location.
%by computing the flux ratios between the observed ratio of the highest S/N recombination line at $\lambda_{\rm{line}}=7.46~\mathrm{\mu m}$ and intrinsic estimates of the line derived from radio observations. This line, the Pfund series 6--5 recombination line, is detected in all pixels with $S/N>10$ and frequently shows $S/N \sim 100$.
%
%To obtain spatially resolved, relative extinction maps between different hydrogen recombination we compare the observed, pixel-wise line ratios with the predicted, intrinsic line ratios.
%
All intrinsic line ratios came from \citet{Storey1995} and are based on Case-B hydrogen recombination emission \citep{Baker1938} for a plasma electron temperature $T_e=7500~\mathrm{K}$ and electron density $n_e=10^3~\mathrm{cm^{-3}}$.
The choice of plasma parameters is unimportant because the relative line ratios depend only weakly on $T_e$ and $n_e$.
\citep[For example, the extremal choice of $n_e=10^6 ~\mathrm{cm^{-3}}$ leads to a percent level error;][]{Storey1995, Scoville2003, Fritz2011}. 
%{\bf The ratio of radio to MIR is almost linearly dependent on $T_e$, though. Is this worth mentioning?} 
% SvF: We have that in equation (2), and discuss that in that section. I'm inclinded to leave it there?
\autoref{tab:hyrdogen_lines} demonstrates that the expected line fluxes can be different by a factor as large as $10$, and the strongest lines are most useful for the creation of spatial extinction maps. 

The line S/N is typically not limited by the photometric sensitivity, the intrinsic line strength, or the extinction. Instead the noise is dominated by systematic errors and in particular residual fringing in the spectrum. 
Highly structured fields like the Galactic center are particularly affected by fringing, and most lines are severely affected. 
%This most severely affects the $\lambda_{\rm{line}}=6.292~\mathrm{\mu m}$, $6.772~\mathrm{\mu m}$,  $8.76~\mathrm{\mu m}$, $11.309~\mathrm{\mu m}$, $12.372~\mathrm{\mu m}$, $16.204~\mathrm{\mu m}$, $16.881~\mathrm{\mu m}$, and the $19.057~\mathrm{\mu m}$ lines.
%
%In contrast, the $\lambda_{\rm{line}}=5.129~\mathrm{\mu m}$, $5.908~\mathrm{\mu m}$, $7.46~\mathrm{\mu m}$, $7.502~\mathrm{\mu m}$ 
The lines that have sufficiently high $S/N$ or can be corrected well enough for fringing to be negligible are marked in Table~\ref{tab:hyrdogen_lines}.

Currently, there is no universally accepted strategy for de-fringing the spectra. We started with the default de-fringing tools provided by the JWST pipeline and then applied a three-step technique to reduce the fringing further and to assess the systematic uncertainty.
In the first step, we used factor analysis (FA)\footnote{Factor analysis is similar to principal component analysis but models basis functions based on the co-variance of the data instead of the variance. We used the factor analysis implemented in \textsc{scikit-learn}.} to derive basis functions that describe the fringing profile locally for each line. For this step, we masked out the line itself for $\pm 600~{\kms}$ around the line center to avoid the basis function being contaminated by the line.
The next step used a Gaussian process (GP, implemented in \textsc{george}; \citealt{Ambikasaran2015_george}) with a mixture of a periodic kernel and non-periodic kernel to model the fringes in the cut-out spectrum, and the final step 
was to interpolate the derived mean posterior GP across the masked-out region. 

\begin{figure*}
    \centering
    \includegraphics[width=0.885\textwidth]{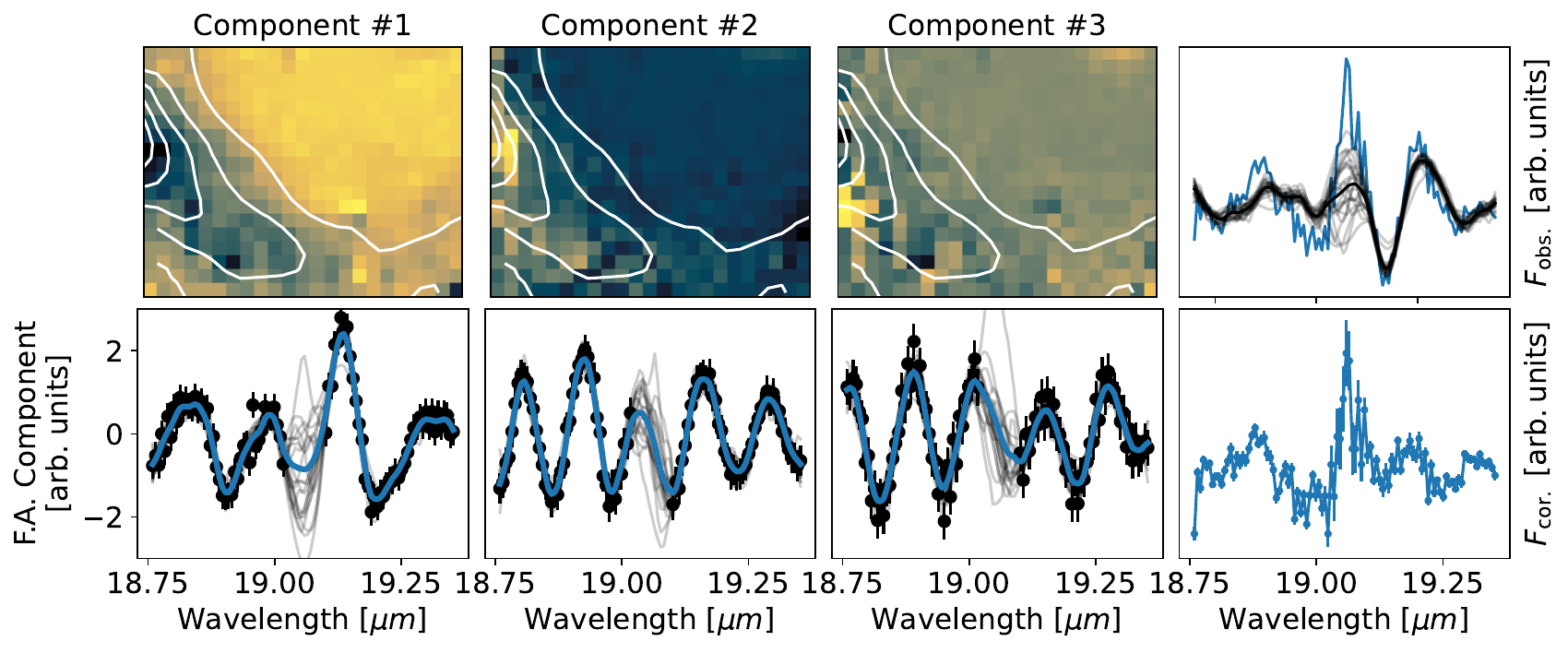}
    \caption{Example of  defringing using the factor-analysis approach for the $\lambda=19.057~\mathrm{\mu m}$ recombination line. The left three plots in the top row show the FA component fields with contour lines showing the channel~4 mean flux density. The left three plots in the bottom row show the respective FA basis functions, the mean posterior GP fit (blue line), and the posterior samples (grey lines). The right-most two plots show the observed spectrum with the derived continuum model (top panel, black and grey lines show continuum models) and the corrected spectrum (bottom panel).}
    \label{fig:defringing}
\end{figure*}

\autoref{fig:defringing} illustrates the defringing method for the $\lambda_{\rm{line}}=19.057~\mathrm{\mu m}$ recombination line.
The amplitudes of the FA basis function (top/left three panels) show clear correlations with the total intensity,  illustrating the dependence of the fringing on the field properties. 
%The first three basis functions are shown in the bottom/left three panels. The region around the line is masked, and the blue line shows the mean posterior GP, the light grey lines shows posterior samples. The uncorrected spectrum together with derived fringing solution is shown in the top/right panel, the residual below, the error bars show the uncertainty derived from the posterior samples.
%
At minimum, this approach quantifies the systematic uncertainty introduced by the fringing and its correction. In the best case, the fringing signal can be subtracted. 
\autoref{fig:Line_measurements} shows the line measurements after de-fringing for two selected apertures centered on the \Ms\ and \Sg.

\begin{table}[]
    \centering
\caption{Hydrogen recombination lines}
    \begin{tabular}{lrc}
\hline\hline
     $~~\lambda$& Transition& Line emissivity\\ $\mathrm{[\mu m]}$&&$10^{-28}$ erg s$^{-1}$ cm$^{-3}$\\
    \hline
    \05.129\rlap{$^{\ast\dag}$}& 10--6 & 5.892\\
    \05.908\rlap{$^\ast$}& \09--6 & 8.199\\
    \06.292& 13--7 & 1.419\\
    \06.772& 12--7 & 1.783\\
    \07.460\rlap{$^\ast$}& \06--5 & 44.91\0\0\\
    \07.502\rlap{$^\ast$}& \08--6 & 11.83\0\0\\
    \07.508& 11--7 & 2.99\0\\
    \08.760\rlap{$^\dag$}& 10--7 & 4.015\\
    10.503& 12--8 & \tablenotemark{a}\\
    11.309& \09--7 & 5.526\\
    12.372\rlap{$^\dag$}& 7--6 & 17.28\0\0\\
    % 12.387\rlap{$^\dag$} & 11--8 & b\\
    16.204& 10--8 & 2.802\\
    16.881& 12--9 & 1.183\\
    19.057\rlap{$^\dag$}& \08--7 & 7.55\0\\
         \hline
    \end{tabular}
\raggedright
\null\\[1ex]
{$^\ast$}Lines with negligible fringing effects.\\
{$^\dag$}Lines included in the staring observations.
\tablenotetext{a}{This line is not usable because it is  blended with the much stronger [\ion{S}{4}] line at 10.510~\micron.}
\tablecomments{Lines detected at $S/N>6$ are listed.  Columns show wavelength, upper and lower principal quantum numbers, and line emissivity for a $n_e=10^3$, $T=7500~\mathrm{K}$ plasma based on the corresponding table by \cite{Storey1995}.}
    \label{tab:hyrdogen_lines}
\end{table}

\subsection{Absolute Line-derived Extinction}
\label{s:abs}
Because the emission at wavelengths longer than the sub-mm  is not affected by extinction \citep[e.g.,][]{Draine2001}, the absolute extinction of any MIR recombination line can be computed by comparing {the observed line flux to the  line flux calculated from radio free-free emission}.  
The present work used the $15~\mathrm{GHz}$ free-free emission to calculate the intrinsic flux of the $7.46~\mathrm{\mu m}$.
This assumes that all of the radio-continuum emission is free-free {and that Case~B recombination theory applies.}

%In order to obtain absolute extinction values for each line, we calibrated each relative line against an estimate of the absolute extinction for the $\lambda=7.46~\mu$m line.
%
%To obtain the absolute value for this line, we followed the procedure detailed by \cite{Scoville2003} and \cite{Fritz2011}:

At 15~GHz, the \Ms\ should be dominated by free-free emission \citep[e.g.,][]{Zhao2003}, and therefore we measured the 7.460~\micron\ line flux in an aperture centered on the \Ms. We fit the line with a triple Gaussian  and obtained an integrated line flux of $f_{7.46~\mathrm{\mu m}} = (4.26\pm0.41)\times 10^{-14}~\mathrm{erg ~cm^{-2}~s^{-1}}$.

% \begin{figure}
%     \centering
%     \includegraphics[width=0.485\textwidth]{figures/lineplot.png}
%     \caption{Line flux of hydrogen recombination lines extract for an aperture centered on the mini-spiral in the MIR JWST and sub-mm SMA and ALMA observations.}
%     \label{fig:746_line}
% \end{figure}

% \subsubsection{H$42\alpha$ estimate}\label{sec:h42}
% To obtain the H$42\alpha$ to $\lambda=7.46~\mathrm{\mu m}$ line ratio, we obtain the integrated line flux from the ALMA line maps.We obtain an integrated line flux of $f_{H42\alpha}=(3.487\pm0.333)\times10^{-19}~\mathrm{erg ~cm^{-2}~s^{-2}~Hz^{-1}}$. 

% The expected line ratios is $r_{\rm{7.46/H42\alpha}} = 205068.5$, from which derive an absolute extinction of $A_{\rm{7.46~\mu m}} = 0.56\pm0.14$.

% We caution that the H$42\alpha$ line shows signs of being optically thick, i.e., a P-cgny like line profile (see \autoref{fig:746_line}). Because of that, the $H42\alpha$ may be underestimated which would imply higher extinction values.

In order to estimate the radio flux density from the same aperture, we used \texttt{reproject} to rescale the VLA images to the JWST pixel scale. The maps were aligned using the MIR and radio bright source IRS~21, which is close to the \Ms, thereby minimizing the impact of any astrometric distortions in the MIRI data. Still, the astrometric alignment could be uncertain by $\approx$0.5~pixels. To estimate the effect of this uncertainty, we shifted the VLA maps by this amount in all directions and measured the integrated radio flux density in the shifted maps. The result was $F_{\rm{15 GHz}} = (6.09\pm  0.15) ~\mathrm{mJy}$, where the uncertainty is from the alignment.

The intrinsic $\lambda=7.46~\mathrm{\mu m}$ line flux is given by \citep{Scoville2003}:
\begin{equation}
\begin{split}
    f_{\rm{7.46}} =& 
    1.5410^{-13} \times \left(\dfrac{T_e}{6000~\mathrm{[K]}}\right)^{-0.52}  \times \left( \dfrac{\nu}{5~\mathrm{[GHz]}}\right)^{0.1} \\
    &\times r_{\rm{P\alpha}} ~\mathrm{erg ~ s^{-1} ~ cm^{-2} ~ mJy^{-1}}\quad, \label{eq:ff_emission}
\end{split}
\end{equation}
where the factor $r_{\rm{P\alpha}}=0.078$ is the line flux ratio between the $P\alpha$ and the $7.46~\mathrm{\mu m}$ line \citep{Storey1995}.
% \end{multicols}
%
The  electron temperature needs to be estimated independently.  \citet{Fritz2011}  obtained  $T_e=(6800\pm500)~{\mathrm{K}}$, and using the same value makes our measurements---$F_{\rm{15 GHz}}$,  $f_{\rm{7.46, exp}}=(7.669\pm0.189)~\mathrm{erg~ s^{-1}~cm^{-2}}$ and  $A_{7.46~\micron}=(0.64\pm0.11)~\mathrm{mag}$---directly comparable to theirs ($A_{\lambda=7.46\mathrm{\mu m;~Fritz~+~  2011}} = (0.81\pm0.23)~\mathrm{mag}$). Differing values of $T_e$ can directly be scaled using \autoref{eq:ff_emission}.
Extinction in other lines shown in \autoref{fig:Line_measurements} can be derived by scaling their relative extinction to this absolute value \citep[cf.\null][]{Scoville2003,Fritz2011}.

\subsection{Spatial Dependence of Line-derived Extinction}
The relative extinction between any pair of spatial positions can be measured from the ratios of hydrogen recombination lines according to the recombination theory described in Section~\ref{s:abs}.  In practice, ratios {using} the strong 7.46~\micron\ line are most useful.

To obtain the line maps, we fit a six-parameter double-Gaussian\footnote{We chose a single Gaussian model when its $\chi^2$ was lower.} line profile to each pixel's spectrum.{Double Gaussian profiles are required because most regions near \Sg, including the \Ms, have two velocity components \citep[e.g.,][]{Nitschai2020}}. The line flux was obtained by integrating the line model $F_{\lambda,\rm model}$ {over a range of} $\pm 1500 ~\mathrm{km~s^{-1}}$ around the peak of the line. 
To identify and remove bad pixels, we kept only pixels with $S/N$ above a minimum value. For the $7.46~\mu$m line, the requirement was $S/N>8$, while for most other lines, the requirement was $S/N>6$. For the weak 
$\lambda=6.772~\mathrm{\mu m}$ line and the three longest-wavelength lines, we accepted $S/N>3$ to allow at least some detections. 

Four recombination lines have circumstances that make it impossible to compute accurate relative-extinction maps.
While the 8--6 line maps are of high quality, the line is blended with the 11--7 line, and neither line is usable. The 12--8 recombination line  is blended with  [\ion{S}{4}]. Finally, fringing for the 10--7 line is too strong to allow for credible line measurements, even after our correction attempt.

\begin{figure*}[htbp] % Use figure* for two-column wide figures
    \centering
    \begin{subfigure}[b]{0.48\textwidth}
        \includegraphics[width=\textwidth]{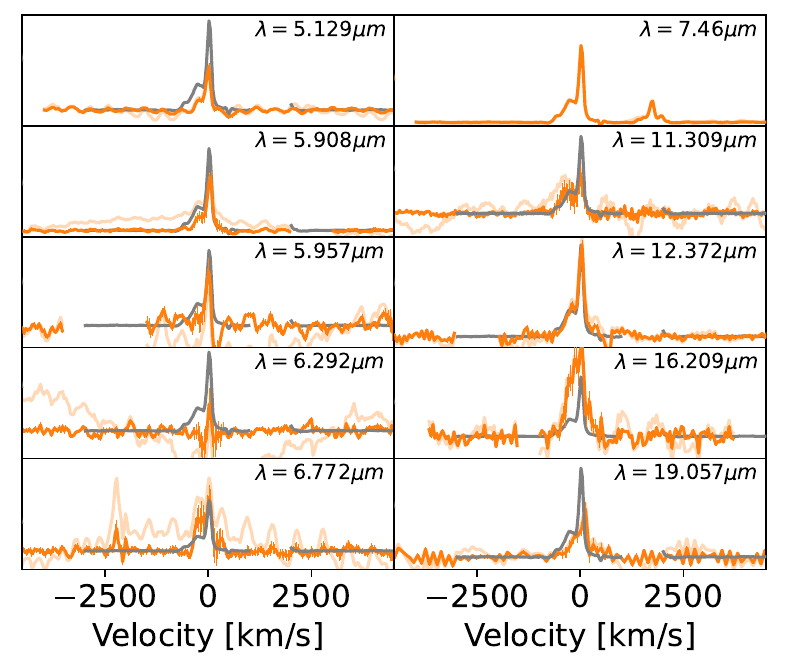}
    \end{subfigure}
    \hfill
    \begin{subfigure}[b]{0.48\textwidth}
        \includegraphics[width=\textwidth]{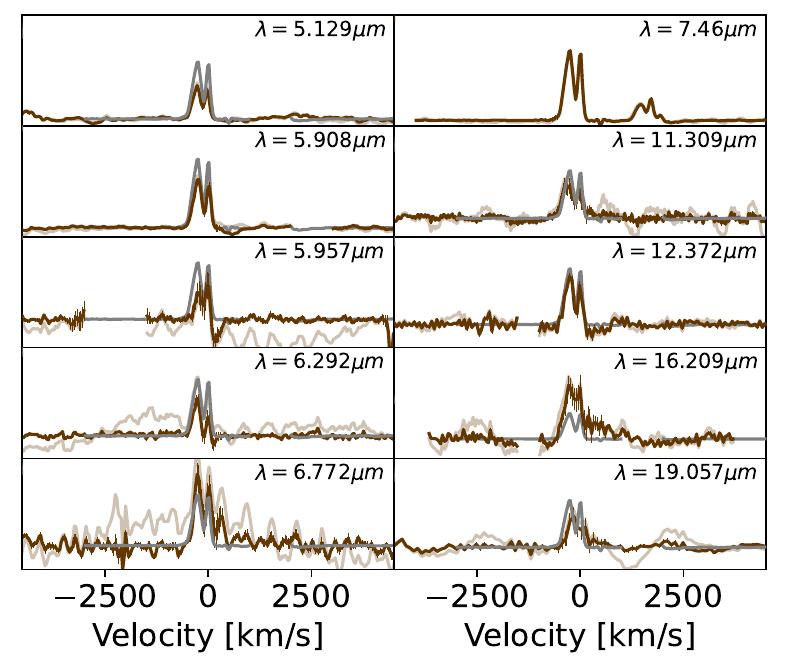}
    \end{subfigure}
    \caption{Hydrogen recombination line measurements for two apertures, centered on \Sg\ (left) and the \Ms\ (right). {The colored lines show the observed spectra after continuum subtraction and fringing correction, and the semi-transparent lines show the spectra before the residual fringing correction.} Each plot shows one of the sufficient-S/N lines (orange and brown lines). The grey line shows the $\lambda_{\rm{line}}=7.46~\mathrm{\mu m}$ recombination line {profile} scaled by the expected line ratio as given in \autoref{tab:hyrdogen_lines}. The respective apertures are shown in \autoref{fig:spatial_map_5.129}.}
    \label{fig:Line_measurements}
\end{figure*}

\begin{figure}
    \centering
    \includegraphics[width=0.48\textwidth]{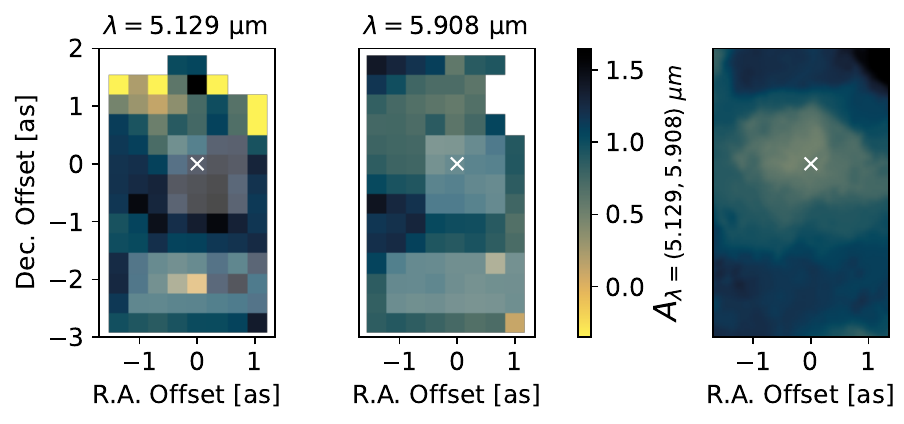}
    \caption{Spatial variation of the line-derived extinction values for the 5.129~\micron\ and 5.908~\micron\ hydrogen recombination lines. The color bar shows the relative-extinction measurements from the line ratios ($A_{5.129/7.46}$ and $A_{5.908/7.46}$) after they were converted to absolute extinction values by adding $A_{7.46}=0.64~\mathrm{mag}$ derived from the \Ms\ extinction measurement (Section~\ref{s:abs}). The white cross shows the approximate location of \Sg, and the lighter patches illustrate the two aperture masks used to obtain the line measurements in \autoref{fig:Line_measurements}. The right-most figure shows the $K$-band extinction map derived from stellar colors \citep{Schodel2010}. The $K$ band has higher  absolute extinction ($\left<A\right>_{2.16}\approx2.4~\mathrm{mag}$) but shows  spatial variations similar to the MIR\null.}
    \label{fig:spatial_map_5.129}
\end{figure}

Based on the pixel-by-pixel integrated line flux measurements, we obtained spatially resolved relative extinction maps, where, in each pixel, the relative extinction to $\lambda_{7.46~\mathrm{\mu m}}$ hydrogen combination line is reported.
For two lines, the 10--6 and the 9--6 lines at  $5.129~\mathrm{\mu m}$ and $5.908~\mathrm{\mu m}$ respectively, the fringing correction is good and the S/N is high. The resulting extinction maps (\autoref{fig:spatial_map_5.129}) show spatial variations on the order of $\Delta A_\lambda\approx0.5$ mag with some individual outlying pixels.
The spatial variations generally agree with the $K$-band maps \citep{Schodel2010}, which have higher angular resolution.\footnote{The absolute amount of $K$-band extinction is more than double the $\sim$5~\micron\  extinction.}
Still, even for these high S/N lines, we cannot rule out that some of the structure in the spatial maps is affected by residual, uncorrected  fringing in the spectra, as can be seen in \autoref{fig:Line_measurements}.
For all other lines, the spatial extinction maps are strongly affected residual fringing artifacts and are  not reliable. (The maps are shown in Appendix~\ref{app:spatial_maps}, \autoref{fig:all_spatial_lines}.)

Following \cite{Fritz2011}, we obtained relative extinction measurements per line by averaging all pixel-by-pixel variations. Absolute extinction measurements were obtained by adding $A_{7.46\micron}$. \autoref{tab:relative_extinction} reports the median extinction values derived, where the reported RMS is the total scatter, which for high $S/N$ lines can be affected by the spatial variations.  The violin plots in \autoref{fig:combined_extinction_law} illustrate the spread of the relative extinction measurements. 

An alternative approach to using pixel-by-pixel line flux measurements is to use apertures centered on specific regions. This approach has the advantage that at least some of the residual fringing in the spectrum is averaged.
We obtained line flux measurements for two apertures centered on the \Ms\ and \Sg\ respectively. The line measurements are shown in \autoref{fig:Line_measurements} and the apertures in \autoref{fig:spatial_map_5.129}. 
%The resulting extinction measurements are shown as brown and orange points \autoref{fig:combined_extinction_law}. 
Both methods generally agree, but the observed scatter also demonstrates that the residual fringing in the lines affects the measurements systematically and limit this approach.

% \begin{figure*}
%     \centering
%     \includegraphics[width=0.985\textwidth]{figures/spatial_extinction_maps_withFlux.png}
%     \caption{Three left most panels: spatial extinction maps based on the line ratios between the $7.46~\mathrm{\mu m}$ and other MIR-hydrogen recombination lines. We've excluded the $6.292~\mathrm{\mu m}$, $6.881~\mathrm{\mu m}$, and $19.057~\mathrm{\mu m}$ recombination line because the spatial structure is dominated by residual uncertainty. Similarly, the $19.057$ recombination line is dominated by systematic scatter and should be interpreted cautiously. The top and bottom panels have the same minimum and maximum color bar range, but differ respectively because of the much higher extinction values at longer wavelengths.
%     Two right most panels: average extinction law plotted along side spectra of the three pixel highlight in the left panels.}
%     \label{fig:spatial_extinctionmaps}
% \end{figure*}

\begin{table}[]
    \centering
\caption{Line-derived Extinction Values toward the \Ms}
\begin{tabular}{lcc}
\hline\hline
    Line & Extinction & MAD $/\sqrt{n}$\\
    \hline
         $\05.129$ & $1.1^{+0.4}_{-0.3}$ & $0.03$\\
         $\05.908$ & $0.9^{+0.1}_{-0.2}$ & $0.01$\\
         $\05.957$ & $1.2^{+0.5}_{-0.5}$ & $0.08$\\
\07.460 & $0.64\pm0.11$\rlap{\tablenotemark{a}}\\
         $\06.292$ & $1.3^{+0.5}_{-0.7}$ & $0.08$\\
         $11.309$& $1.3^{+0.3}_{-0.6}$ & $0.03$\\
         $12.372$& $1.4^{+0.5}_{-0.6}$ & $0.03$\\
%12.387\\
         $16.204$& $0.3^{+0.5}_{-0.6}$ & $0.05$\\
         $19.057$& $2.0^{+0.5}_{-0.7}$ & $0.07$\\
\hline
\end{tabular}
\raggedright
\tablenotetext{a}{absolute extinction from Section~\ref{s:abs}.}
\tablecomments{Relative values are scaled to the 7.46~\micron\ hydrogen 6--5 recombination line ($A_{7.46\micron}=0.60\pm0.18$). 
Quoted uncertainties report the $16\%$- to $84\%$-percentile range.
The last column shows the standard error based on the median absolute deviation (MAD) value (scaled by $\times 0.67449$ to obtain the $1 \sigma$ confidence range). 
The standard error is underestimated due to non-Gaussianity caused by residual systematics in the data.}
    \label{tab:relative_extinction}
\end{table}

% \begin{figure}
%     \centering
%     \includegraphics[width=0.5\textwidth]{figures/line_derived_extinction.pdf}
%     \caption{Line-derived extinction measurements based on hydrogen recombination lines reported in \autoref{tab:relative_extinction}. The relative measurements were calibrated using $A_{7.46}=0.64~\mathrm{mag}$. Violin plots show the spread of values using the pixel-by-pixel flux ratios; the orange and brown points show the measurements for two apertures centered on \Sg\ and the \Ms\ shown in \autoref{fig:Line_measurements}.}
%     \label{fig:line_derived}
% \end{figure}

%\section{Mid Infrared Extinction law for the Galactic center}
\section{Discussion and Summary}
\label{s:sum}
%In order to derive an extinction law that is valid for the Galactic center, we combined the three measurements. First we combined 
The continuum extinction curves  for \Sg\ and the \Ms\ agree with each other. IRS 29N has higher extinction, some of which may arise in the  source itself. The average \Sg+\Ms\ curve agrees well with {extinction} measurements {in four local luminous infrared galaxies} by \citet{Donnan2024}, including their finding of smaller extinction in the 19~\micron\ silicate feature than some previous studies had reported. However, {their} curve's shape depends on the unknown emissivity of the emitting dust, and the absolute extinction was undetermined.

Modeling the MIR dust emission suggests that the extinction is {spatially} variable and is higher in the immediate environment of dust-enshrouded stars like IRS~29. At the same time, we do not see significant variation in the underlying optical depth of the dust (\autoref{fig:opacity}) nor in the dust distribution $\Psi(T,\tau_{9.8})$ (\autoref{fig:dust_properties}). 
This implies that the attenuating extinction is dominated by a single, absorbing screen. This is consistent with results in other studies \citep[e.g.,][]{Fritz2011} and the screen typically thought to be in situated in the Central Molecular Zone (CMZ, e.g.,  \citealt{Morris1996,Launhardt2002,Schodel2010,Nogueras-Lara2018}). 

The absence of variability in the opacity profile (\autoref{fig:opacity}) implies that, while the extinction magnitude in the CMZ is locally variable, the dust properties giving rise to that profile are similar. Intriguingly, the deviation from the \cite{Kemper2004} law found for several AGN tori (e.g., VV114 NE\&SW, NGC 356 S, NGC 7456 Nuc) has a similar shape albeit a difference in the ratio of the $9.8~\mathrm{\mu m}$-to-$18~\mathrm{\mu m}$ silicate absorption features. {This similarity indicates that the opacity profile may hold generally for MIR dust spectra as these extragalactic AGN tori probe very different scales and  host-galaxy properties.}

The hydrogen recombination-line ratios give relative extinctions but only at the wavelengths of the lines.  The line extinctions are generally consistent with the derived continuum extinction within the uncertainties, but there are no useful lines near the peak of either silicate feature. For the lines that exist, fringing is a major uncertainty on the measured line flux and therefore on the extinction. 

\begin{figure}
    \centering
    \includegraphics[width=0.5\textwidth]{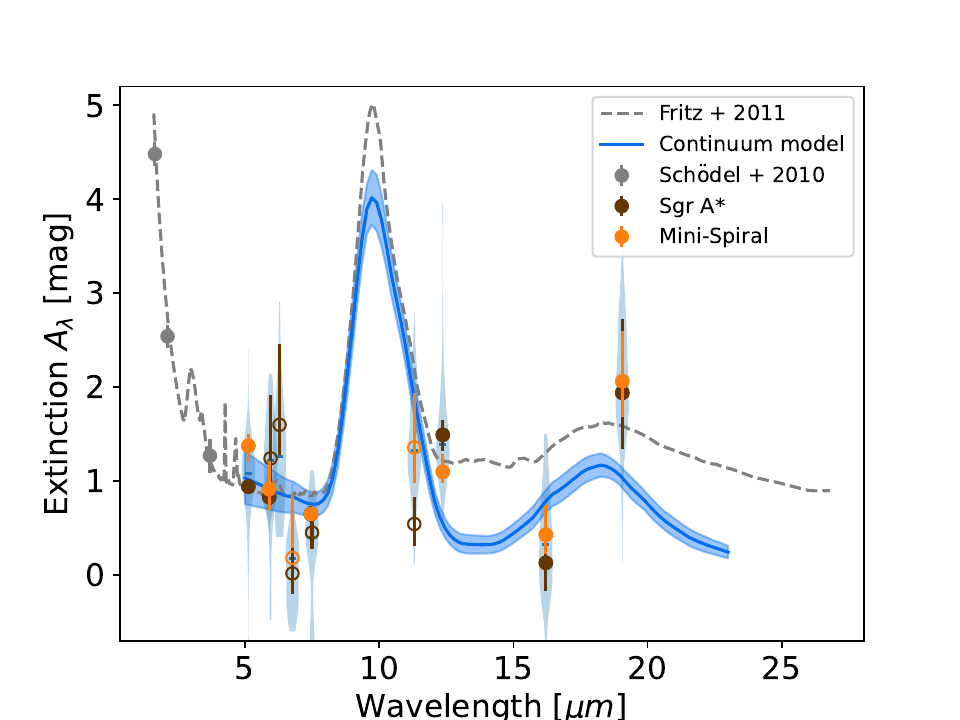}
    \caption{Best overall MIR extinction law. The blue curve shows the continuum extinction  
    %law is based on the dust continuum model 
    shifted by  $-0.18\pm0.02~\mathrm{mag}$ to match the line-derived extinction values ({filled brown and orange} points). The continuum model is the average of the Sgr A* and \Ms\ model. The fit excludes the four lines most affected by residual fringing: 5.957, 6.292, 6.772, and 11.309~\micron, indicated by open points. The grey dashed line shows the extinction law derived by \cite{Fritz2011}, and grey points show the stellar-color-based extinction law by \cite{Schodel2010}. 
    The relative extinction measurements were calibrated using $A_{7.46}=0.64~\mathrm{mag}$. Violin plots show the spread of values using the pixel-by-pixel extinction measurements; the {brown and orange} points show the measurements for two apertures centered on \Sg\ and the \Ms\ shown in \autoref{fig:Line_measurements}.
    %In this fit, we excluded those lines most affected by residual fringing: $\lambda_{\rm{line}}=5.957~\mathrm{\mu m},6.292~\mathrm{\mu m},6.772~\mathrm{\mu m},11.309~\mathrm{\mu m}$. The best fit is shifted by $\Delta A_{\lambda} = -0.18\pm0.02$.
    }
    \label{fig:combined_extinction_law}
\end{figure}

The strongest recombination line, $n=6$--5, gives an absolute extinction measurement at $7.46~\mathrm{\mu m}$.  The fringing correction is well determined, but any 15~GHz emission that is not free-free would mean the
true extinction is smaller than the measured value.
Taking the uncertainties into account, Figure~\ref{fig:combined_extinction_law} gives the best estimate for a combined extinction curve. 
%The resulting best fit continuum model, our best-guess continuum model, is shown in \autoref{fig:combined_extinction_law}.
The grey extinction needed to match the average of the lines is $-0.18\pm0.02~\mathrm{mag}$, i.e., slightly less extinction than was built in to the continuum curve.

Our extinction measurements agree with the measurements by \cite{Fritz2011} for wavelengths shorter than $\sim$8~\micron. At the peak of the first silicate absorption feature, \cite{Fritz2011} reported  $A_{9.8\micron}\sim 5.5$ mag, which is higher than we found for the \Ms\ and \Sg. At the same time, {their} value agrees with the extinction measured for IRS 29N\null. It is plausible that the difference stems from the much larger field of view of ISO/SWS and the spatial variation seen in the extinction. At wavelengths longer than $\sim$13~\micron, our extinction measurements are inconsistent with the measurements by \cite{Fritz2011}. Those authors did not model the dust continuum directly, and the difference most likely stems from the interpolation that \cite{Fritz2011} used, which they cautioned is uncertain because of the limited number of lines at longer wavelengths.

\begin{figure}
    \centering
    \includegraphics[width=0.485\textwidth]{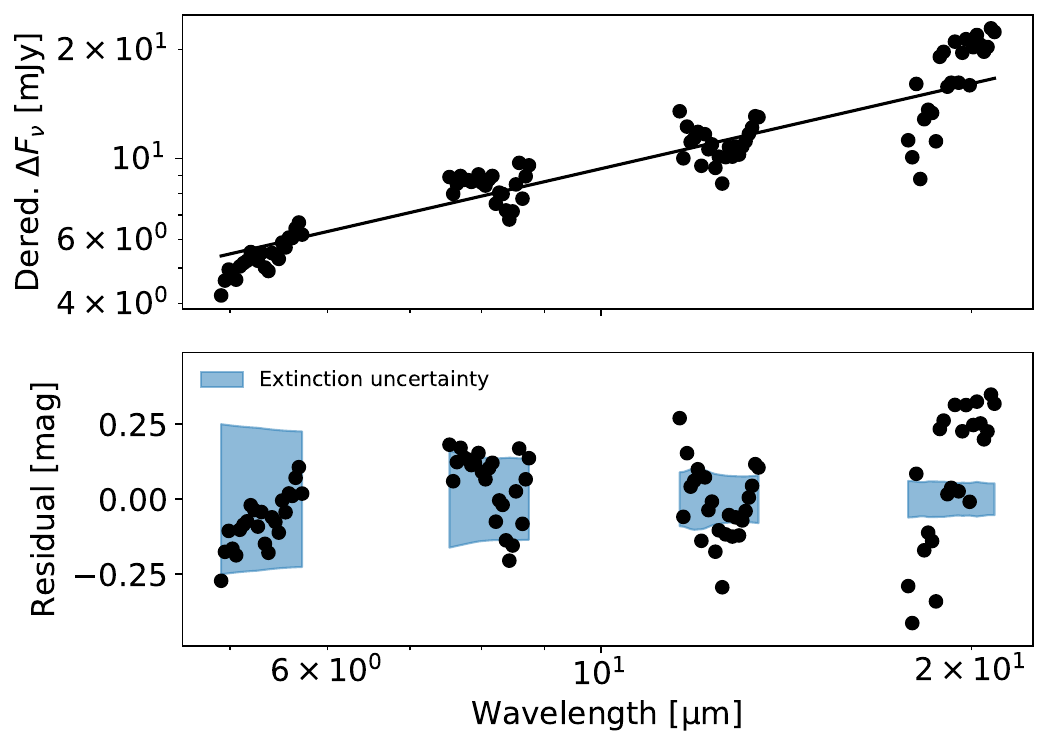}
    \caption{Extinction correction applied to a bright state of \Sg. The top panel shows a power-law fit to the differential flux {measurements}, and the bottom panel shows the deviations from a power law. The blue areas indicate the extinction uncertainty derived from the dust-continuum modeling. The data were first presented by \cite{vonFellenberg2025}, and \cite{Michail_SED} gave a detailed analysis of the spectrum.}
    \label{fig:sgra_spec}
\end{figure}

A companion paper \citep{Michail_SED} reports the SED of a \Sg\ flare. 
The observations were among the staring ones used here and included only the four short sub-bands of the MRS\null.
The SED was derived on the assumption that the emission in the four channels is synchrotron and therefore has a power-law SED \citep[e.g.,][]{GravityCollaboration2021_xrayflare}. The same SED assumption gives an estimate of the residual uncertainty of the extinction calibration. We find deviations from a power law on the order of 0.2~{mag} for Ch1, Ch2, and Ch3 and $\sim$0.3~{mag} for Ch4. These deviations are consistent with the extinction error bars derived from the continuum modeling (\autoref{fig:sgra_spec}).

%\section{Conclusion}
%In this paper, we derived the first mid infrared extinction law for the Galactic center at JWST-spatial scales. We used three different approaches to derive dust-continuum based extinction values; line-ratio based spatially-resolved, relative extinction values; and a absolute line-derived extinction calibration for the high $S/N$ $\lambda=7.46~\mathrm{\mu m}$ (Pfund 6--5) hydrogen recombination line.

All measurements are affected by systematic uncertainties. The dust-continuum-based extinction law is inherently degenerate with the strength and shape of the dust emissivity. The emission-line maps suffer from residual fringing and undersampling of the PSF\null. Finally, the absolute line-derived extinction calibration is subject to uncertainty from the unknown electron temperature and whether the 15~GHz flux is entirely from free-free emission.

We leveraged these different systematics by combining the measurements and obtained a  ``best-guess'' extinction law for the MIR\null. By applying the extinction law to the flare measurements reported by \cite{Michail_SED}, we estimate that this ``best-guess'' law is uncertain by approximately {0.2 mag to 0.3 mag} depending on the channel, a value that allows for predictive measurements of Galactic-center mid-infrared sources.

%% The "ht!" tells LaTeX to put the figure "here" first, at the "top" next
%% and to override the normal way of calculating a float position.
%% The asterisk after "figure" tells the compiler to span multiple columns
%% if a two column style is selected.

%% Please use the acknowledgment and contribution environments. This will 
%% be anonomyized when the "anonymous" style option is used. 
\begin{acknowledgments}
We thank Jongseo Kim and Julian R{\"u}stig for their help and comments on the implementation of the dust continuum model with \textsc{nifty}.

This research was supported by the International Space Science Institute (ISSI) in Bern, through ISSI International Team project \#24-610, and we thank the ISSI team for their generous hospitality.

SDvF gratefully acknowledges the support of the Alexander von Humboldt Foundation through a Feodor Lynen Fellowship and thanks CITA for their hospitality and collaboration.

SDvF, BR, BSG are supported by the Natural Sciences \& Engineering Research Council of Canada (NSERC), the Canadian Space Agency (23JWGO2A01), and by a grant from the Simons Foundation (MP-SCMPS-00001470). BR acknowledges a guest researcher position at the Flatiron Institute, supported by the Simons Foundation.

JM is supported by an NSF Astronomy and Astrophysics Postdoctoral Fellowship under award AST-2401752. 

DH, ZS, NMF acknowledge support from the Canadian Space Agency (23JWGO2A01), the Natural Sciences and Engineering Research Council of Canada (NSERC) Discovery Grant program, the Canada Research Chairs (CRC) program, the Fondes de Recherche Nature et Technologies (FRQNT) Centre de recherche en astrophysique du Québec, and the Trottier Space Institute at McGill.
NMF acknowledges funding from the FRQNT Doctoral Research Scholarship.

TR acknowledges funding support from the Deutscher Akademischer Austauschdienst (DAAD) Working Internships in Science and Engineering (WISE) program.

The National Radio Astronomy Observatory is a facility of the National Science Foundation operated under cooperative agreement by Associated Universities, Inc. 

% This paper makes use of the following ALMA data: ADS/JAO.ALMA\#2023.1.00295.S, ADS/JAO.ALMA\#2018.A.00052.S. ALMA is a partnership of ESO (representing its member states), NSF (USA) and NINS (Japan), together with NRC (Canada), NSTC and ASIAA (Taiwan), and KASI (Republic of Korea), in cooperation with the Republic of Chile. The Joint ALMA Observatory is operated by ESO, AUI/NRAO and NAOJ. 

This work is based [in part] on observations made with the NASA/ESA/CSA James Webb Space Telescope. 
The data were obtained from the Mikulski Archive for Space Telescopes at the Space Telescope Science Institute, which is operated by the Association of Universities for Research in Astronomy, Inc., under NASA contract NAS 5-03127 for JWST. These observations are associated with program \#4572.
The observations are available at the Mikulski Archive for Space Telescopes (\url{https://mast.stsci.edu/}).
% CfA
Support for program \#4572 was provided by NASA through a grant from the Space Telescope Science Institute, which is operated by the Association of Universities for Research in Astronomy, Inc., under NASA contract NAS 5-03127.

Some of the results in this publication have been derived using the
NIFTy package (https://gitlab.mpcdf.mpg.de/ift/NIFTy)

\end{acknowledgments}

\begin{contribution}
%%This section gives authors the space to recognize author contributions. The text inside this environment is NOT counted towards the total word quanta. At a minimum, manuscripts are expected to include this text:

SDvF contributed to continuum dust model, line derived extinction modeling, and manuscript writing.
JM contributed to data calibration, line derived extinction modeling, and manuscript writing.
SPW contributed to scientific discussion and manuscript review.
BSG, TR, and MGM contributed to data analysis and scientific discussion.
GGF, NMF, DH, JLH, HAS, SZ, and GW contributed to proposal and manuscript writing.

%% But authors are expected to provide more specific details, e.g. 
%%
%%SC was responsible for writing and submitting the manuscript.
%%WWM came up with the initial research concept and edited the manuscript.
%%OTS obtained the funding and edited the manuscript.
%%EBF provided the formal analysis and validation. He also edited the manuscript.
%%GEH Supervised the undergraduates, wrote the software and administers the project github and Zenodo repositories.
%%
%% Authors can use the Contributor Role Taxonomy (CRediT) at
%% for ideas on how write a good statement tailored to their needs.

\end{contribution}

The JWST data presented in this article were obtained from the Mikulski Archive for Space Telescopes (MAST) at the Space Telescope Science Institute. The specific observations analyzed can be accessed via \dataset[doi: 10.17909/6mxr-st18]{https://doi.org/10.17909/6mxr-st18}
\facilities{JWST, VLA}

\software{astropy\citep{astropy:2013,astropy:2018,astropy:2022},  
nifty.re8 \citep{niftyre,Arras2021,Arras2022,Frank2021,knollmüller2020metric}
          }


\appendix

\section{Spatial Extinction maps} \label{app:spatial_maps}
\begin{figure*}
    \centering
    \includegraphics[width=0.985\textwidth]{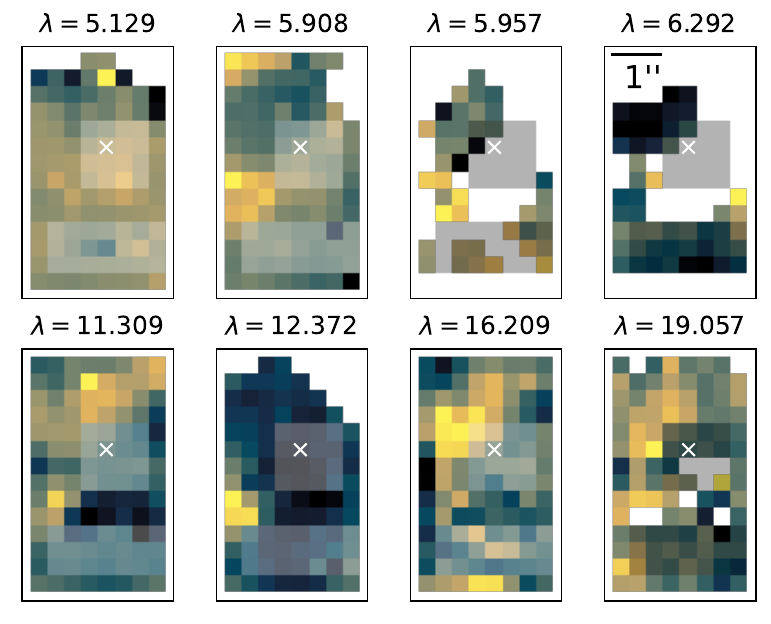}
    \caption{{Spatial variation of the line-derived extinction values for all lines with $\rm{S/N}>6$. Pixels with $\rm{S/N}<6$ are flagged. As in Figure~\ref{fig:spatial_map_5.129}, the white cross marks the \Sg\ position, and the lighter or grey patches indicate the two aperture areas where the lines were measured.} Residual fringing {may have affected} the spatial correlation. (See text for details.) The color maps have different minima and maxima, but have been scaled to illustrate the spatial variations in the field. {The white crosses mark} the approximate location of \Sg.}
    \label{fig:all_spatial_lines}
\end{figure*}

Spatial extinction maps for all lines with sufficient S/N are provided in \autoref{fig:all_spatial_lines}. We caution that residual fringing may affect the spatial correlation, as it depends on the observed field (compare with \autoref{fig:defringing}). 

\bibliography{sample7}{}
\bibliographystyle{aasjournalv7}



\end{document}